\documentclass[floats,floatfix,amssymb,prd,twocolumn,superscriptaddress,nofootinbib,preprintnumbers]{revtex4-1}
		
\usepackage{amssymb,amsmath,verbatim,mathtools,needspace,enumitem,etoolbox,graphicx,physics,microtype,afterpage,bm}
\usepackage[dvipsnames, usenames]{xcolor}
\definecolor{linkcolor}{rgb}{0.0,0.3,0.5}
\usepackage[unicode, colorlinks=true, linkcolor=linkcolor, citecolor=linkcolor, filecolor=linkcolor,urlcolor=linkcolor, pdfusetitle]{hyperref}
\usepackage[all]{hypcap}
\usepackage[T1]{fontenc}
\usepackage[utf8]{inputenc}
\usepackage{tabularx}
\usepackage{float}
\newcommand{\tn}{\textnormal}
\newcommand{\GSSI}{Gran Sasso Science Institute (GSSI), I-67100 L’Aquila, Italy}
\newcommand{\ENS}{Département de Physique, ENS de Lyon, Univ. Claude Bernard, F-69342 Lyon, France}
\newcommand{\GranSasso}{INFN, Laboratori Nazionali del Gran Sasso, I-67100 Assergi, Italy}

\def\lsim{\mathrel{\rlap{\lower4pt\hbox{\hskip0.5pt$\sim$}}
    \raise1pt\hbox{$<$}}}         
\def\gsim{\mathrel{\rlap{\lower4pt\hbox{\hskip0.5pt$\sim$}}
    \raise1pt\hbox{$>$}}}         

\begin{document}
\title{Black holes surrounded by generic dark matter profiles:\\
appearance and gravitational-wave emission}

\author{Enzo Figueiredo}
\affiliation{\ENS}
\affiliation{\GSSI}

\author{Andrea Maselli}
\affiliation{\GSSI}
\affiliation{\GranSasso}

\author{Vitor Cardoso}
\affiliation{Niels Bohr International Academy, Niels Bohr Institute, Blegdamsvej 17, 2100 Copenhagen, Denmark}
\affiliation{CENTRA, Departamento de F\'{\i}sica, Instituto Superior T\'ecnico -- IST, Universidade de Lisboa -- UL,
Avenida Rovisco Pais 1, 1049 Lisboa, Portugal}

\begin{abstract}
We develop a numerical approach to find asymptotically flat 
black hole solutions coupled to anisotropic fluids, described 
by generic density profiles. Our model allows for a variety of 
applications in realistic astrophysical scenarios, and is potentially 
able to describe the geometry of galaxies hosting supermassive 
black holes, dark matter environments and accretion phenomena. 
We apply our framework to a black hole surrounded by different 
families of dark matter profiles, namely the Hernquist, the Navarro-Frenk 
White and the Einasto models. We study the geodesic motion of 
light and of massive particles in such spacetimes. Moreover we 
compute gravitational axial perturbations induced by a small 
secondary on the numerical background, and determine the changes 
in the emitted gravitational wave fluxes compared to the vacuum case.
Our analysis confirms and extend previous studies showing that 
modifications of orbital frequencies and axial fluxes can be described 
in terms of gravitational-redshift, regardless of the halo model.
\end{abstract}

\maketitle

\section{Introduction}

Astrophysical compact sources do not live in vacuum, but rather evolve 
embedded in a complex environment of plasma, electromagnetic fields, 
and dark matter (DM), which is expected to leave detectable imprints on 
the dynamics and emitted gravitational wave (GW) signals of these 
systems~\cite{Yunes:2011ws,Barausse:2014tra,Cardoso:2019rou,Cardoso:2020iji,Derdzinski:2020wlw,Cardoso:2022whc,Zwick:2022dih}.

Modifications to the orbital phase due to environmental effects provide a 
new route to determine the properties of the astrophysical arena where 
binaries evolve. Exploiting such effects can allow to probe a variety of 
fundamental astro-physics models including the formation channels of 
compact binaries \cite{Cardoso:2022nzc,Pan:2021oob}, the distribution 
of baryonic and dark matter surrounding massive objects \cite{DeLuca:2022xlz,Sberna:2022qbn,Speri:2022upm,Kavanagh:2020cfn,Speeney:2022ryg,Macedo:2013qea,Eda:2013gg}, 
as well as the existence of new fundamental fields coupled to the gravity 
sector \cite{Maselli:2021men,Brito:2015oca}. Understanding how matter 
distribution affects the dynamics of coalescing binaries, and characterise 
the imprint left on the GW generation and propagation mechanisms, requires 
fully relativistic solutions that describe black holes (BHs) within a medium.  

So far, however, most studies have mainly worked on Newtonian inspired 
corrections, or computed changes in the leading quadrupolar GW emission 
to assess the relevance of non-vacuum contributions \cite{Babak:2006uv,Destounis:2021mqv}. 
Few calculations that included relativistic corrections to the background 
spacetime or to dynamical friction highlighted the relevance of such terms, 
which also lead in general to larger modifications to the emitted signals \cite{Vicente:2022ivh,Traykova:2021dua,Sadeghian:2013laa,Speeney:2022ryg}. 
Moreover, all the investigations carried out so far to model GW signals, work 
within a post-Newtonian framework, which provides the best approach to 
describe nearly equal mass systems, but looses its validity for asymmetric 
binaries, and in particular for their more extreme configurations, which are 
expected to provide the tightest constraints on environmental 
parameters \cite{Cardoso:2019rou}.

Such studies however highlight a key feature common to the vast majority of 
studies on environmental effects: corrections to the binary evolution tend to 
affect the low frequency inspiral regime, where coalescing binaries accumulate 
tens of hundreds of orbital cycles in the detector bandwidth of next generation 
of detectors. Developing precise models for BH solutions and the emitted GW 
signals is therefore even more pressing, given the accuracy requirements 
that such future observations will demand.

\begin{figure}[h]
    \centering
    \includegraphics[width=9cm]{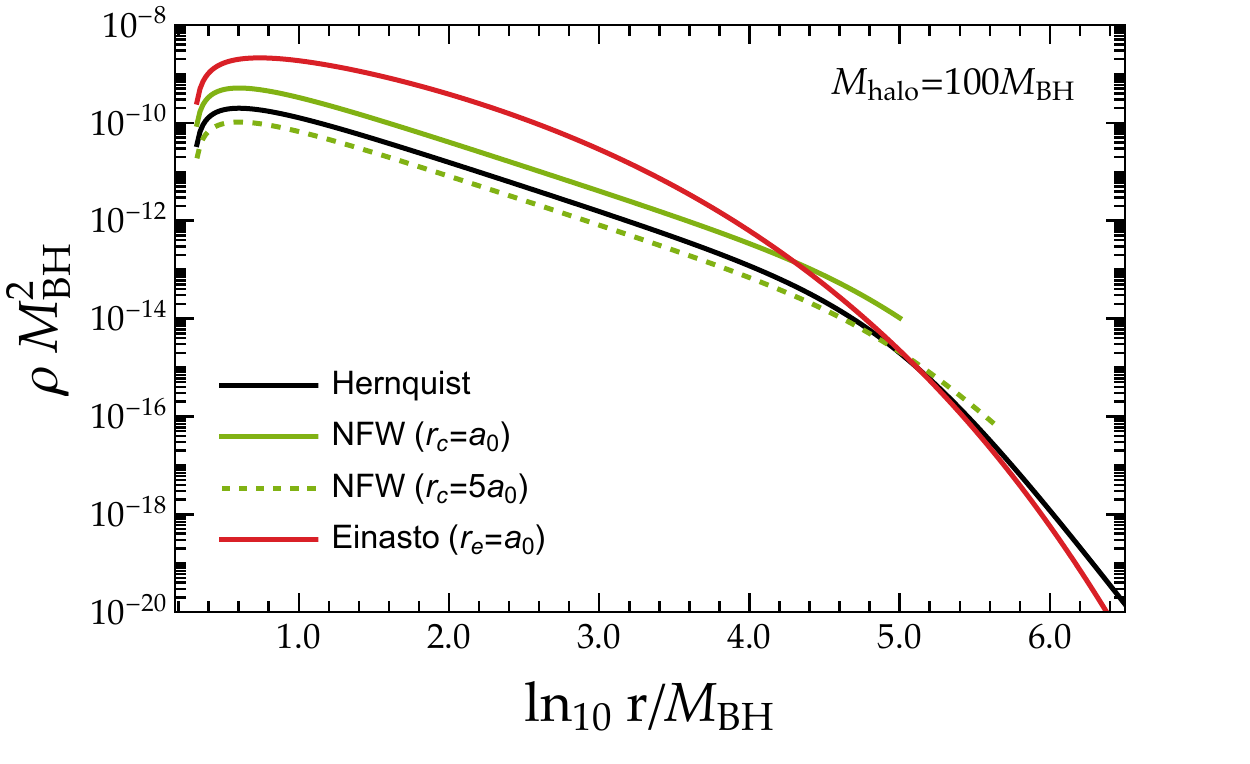}
    \caption{Halo density profiles considered in this work as a function of the 
    distance from the BH (in Schwarzschild-like coordinates), for a specific configuration 
    with $M_{\rm halo}=100M_{\rm BH}$ and $a_0 = 10^5 M_{\rm BH}$.}
    \label{fig:profiles}
\end{figure}
Recently, some of the present authors worked out the first spacetime geometry 
generated by a non-spinning BH within a core of matter~\cite{Cardoso:2020iji}. 
This background was exploited to compute the GW fluxes emitted by extreme 
mass ratio inspirals (EMRIs), taking into account the coupling between fluid and 
gravitational perturbations in the full relativistic theory ~\cite{Cardoso:2021wlq}. 
While these works provide the first setup to study BH physics in realistic dense 
environments, they assumed a specific choice for the matter distribution, given 
by the Hernquist model \cite{Hernquist:1990be}, which allows to compute the 
metric in a closed analytical form.

In this paper we take a step further, and develop a new framework which extends 
the domain of applicability of the solution devised in Ref.~\cite{Cardoso:2020iji} to 
generic matter profiles. We build a numerical pipeline to compute the spacetime 
geometry of spherically symmetric, and asymptotically flat BHs, within a spherically 
symmetric environment. The corresponding matter can be orders of magnitude 
larger than the BH itself, in which case our solution may describe, for example, a 
galactic DM halo. But the geometry could describe some more exotic physics, like 
a BH surrounded by a small-scale matter structure. We apply this formalism to 
investigate the effect of different families  of DM halos on the BH geometry, and its 
geodesic structure. We also compute axial gravitational perturbations induced by a 
point-like on circular motion around the BH, and determine the changes onto the 
GW fluxes in terms of the halo properties. Hereafter we use geometric units 
$G = c = 1$.

\section{Background and axial modes}

In this Section we summarize the key equations that describe a static, spherically 
symmetric BH spacetime embedded in an environment with a generic density profile 
$\rho(r)$. We refer the reader to \cite{Cardoso:2021wlq,Cardoso:2022whc} for an 
extensive discussion, as well for technical details on the formalism, which has been 
originally applied to study binary BHs with an Hernquist-type matter distribution \cite{Hernquist:1990be}. 

We adopt the Einstein cluster approach to model a stationary BH surrounded by 
a collection of gravitating masses \cite{Einstein:1939ms}. In this framework the 
background metric specified by the line element 
\begin{equation}
ds^2=g_{\mu\nu}^{(0)}dx^\mu dx^\nu=-a(r) dt^2 + \frac{d r^2}{1-\frac{2m(r)}{r}} + r^2 d\Omega^2 \ ,
\end{equation}
is a solution of the sourced Einstein's field equations 
\begin{equation}
G^{(0)}_{\mu\nu}=8\pi T^\tn{(0)env}_{\mu\nu}\ ,\label{math:fieldsback}
\end{equation}
where the properties of the environment are encoded by the anisotropic stress-energy 
tensor with the following form:
\begin{equation}
    (T^\tn{(0)env})^{\mu}{_\nu} = diag(-\rho,0,P_t,P_t)\ .
\end{equation}

For a given choice of $\rho(r)$, the mass profile is determined by the continuity 
equation $m'(r)=4\pi r^2 \rho(r)$, while the metric variable $a(r)$ and the tangential 
pressure are determined by the $rr$ component of Eqs.~\eqref{math:fieldsback} and 
by the Bianchi identities, respectively:
\begin{equation}
\frac{a'(r)}{a(r)}=\frac{2m(r)/r}{r-2m(r)}\quad \ ,\quad 
P_t(r) =\frac{m(r)/2}{r-2m(r)}\rho(r)\ ,\label{math:back}
\end{equation}
where a prime denotes a derivative with respect to the radial coordinate.
Equations \eqref{math:back} completely specify the background solution. However, 
unlike the Hernquist profile which allows to compute metric and matter quantities in 
a closed analytical form, hereafter we focus on generic density distributions which 
requires a fully numerical treatment. As we shall describe in the next section, such 
numerical pipeline will only require as starting point a tabulated input for $\rho(r)$. 

The solution for $a(r)$ and $m(r)$ also determines the geodesic properties for 
massive and massless particles. As for the vacuum Schwarzschild counterpart, 
the spacetime admits a timelike and a spacelike Killing vector associated 
with two conserved quantities, which can be identified with the energy per unit 
mass and the specific angular momentum at infinity:
\begin{equation}
E=\left[\frac{r-2m(r)}{r-3m(r)}a(r)\right]^{1/2}_{r=r_p}\  ,
\  
L=\left[\frac{m(r)}{r-3m(r)}\right]^{1/2}_{r=r_p} \ .
\end{equation}
where $r_p$ identifies the particle orbital radius. Geodesics are planar and without 
loss of generality we assume $\theta(r=r_p)=\pi/2$. For massive objects the radius of 
the innermost circular orbit (ISCO) $r_\tn{ISCO}$ is given by a solution of the following 
equation
\begin{equation}
r^2m'(r) + rm(r) - 6m^2(r)=0\ ,
\end{equation}
with the corresponding angular frequency 
\begin{equation}
    \Omega_\tn{ISCO}=\left[\frac{1}{r^2} \frac{a(r) m(r)}{r-2m(r)} \right]^{1/2}_{r=r_{\rm ISCO}}\ .
\end{equation}
Similarly, for massless particles the light-ring is determined by solving the equation $r-3m(r)=0$, 
and its frequency given by 
\begin{equation}
\Omega_\tn{LR}=\frac{\sqrt{a(r_\tn{LR})}}{r_\tn{LR}}\,.
\end{equation}

With the background solution in hand we can study how GW propagation and generation 
change due to the environment. We focus on astrophysical scenarios provided by 
EMRIs, in which a secondary stellar mass object orbits the primary BH inducing 
perturbations of the metric and the stress-energy tensor
\begin{equation}
g_{\mu\nu}=g_{\mu\nu}^{(0)}+g_{\mu\nu}^{(1)}\quad \ ,\quad 
T^\tn{env}_{\mu\nu}=T^\tn{(0)env}_{\mu\nu}+T^\tn{(1)env}_{\mu\nu} \ .
\end{equation}
The first order terms $g_{\mu\nu}^{(1)}$ and $T^\tn{(1)env}_{\mu\nu}$ 
are decomposed into standard axial and polar 
modes \cite{Regge:1957td,Zerilli:1970wzz,Lindblom:1983ps,1967ApJ...149..591T} 
and satisfy the perturbed field equations 
\begin{equation}
G^{(1)}_{\mu\nu}=8\pi T^\tn{(1)env}_{\mu\nu}+8\pi T^p_{\mu\nu}\ ,\label{math:fieldspert}    
\end{equation}
where $T^p_{\mu\nu}$ is the stress-energy tensor associated to the secondary binary 
component with mass $m_p$. Hereafter we focus on axial type perturbations only, 
referring to the polar sector for a future study. In this case, metric fluctuations decouple 
from matter variables and can be cast into a single Schrodinger-like equation for the 
master function 
$\psi_{\ell m}(r)$:  
\begin{equation}
    \frac{d^2\psi_{\ell m}(r)}{dr_*^2}+ \left[ \omega^2 - V_\ell(r) \right]\psi_{\ell m}(r) =\mathcal{S}_{\ell m} \ , \label{math:masterax}
\end{equation}
where $\ell=2,\ldots \infty$, $m=-\ell\ldots\ell$,  $r^\star$ is the tortoise coordinate defined 
as $dr^*/dr = \left[ a(r) (1-2m(r)/r)  \right]^{-1/2}$ and the scattering potential reads 
\begin{equation}
V_\ell(r) = \frac{a(r)}{r^2}  \left[\ell(\ell+1) - \frac{6m(r)}{r} + m'(r)  
\right] \ .
\end{equation}
For circular orbits the source term $S_{\ell m}$ can be written as: 
\begin{equation}
\mathcal{S}_{\ell m}=8 i \pi \sqrt{2a(r)} \sqrt{1-\frac{2m(r)}{r}} 
\Lambda_\ell r [ a(r) \mathcal{D}_{lm}(r)]'\ ,
\end{equation}
with $\Lambda_\ell= [l(l+1)(l-1)(l+2)]^{-1/2}$ and
\begin{align}
    \mathcal{D}_{l m} =m\Lambda_l m_p &\frac{L^2}{E} \sqrt{\frac{a(r)(r-2m(r))}{2r^9}} \delta(r-r_p)\ \times \nonumber\\
    &\delta(\omega -m \omega_p) \frac{\partial}{\partial\theta}Y_{\ell m}\bigg\vert_{\theta=\pi/2,\phi=0},
\end{align}
where $Y_{\ell m}(\theta,\phi)$ are the standard spherical harmonics, and $\omega_p$ 
is the secondary orbital frequency evaluated at $r=r_p$. For $m(r)=M_\tn{BH}$ 
Eq.~\eqref{math:masterax} reduces to the well-known Regge-Wheeler equation.

\subsection*{The environmental density profiles}
We consider parametric density distributions described by the following 
semi-analytic relation:
\begin{equation}
\rho(r)=\rho_0 (r/a_0)^{-\gamma}[1+(r/a_0)^\alpha]^{(\gamma-\beta)/\alpha}\ .
\label{math:densityp}
\end{equation}
For a given choice of the coefficients $(\alpha,\beta,\gamma)$, Eq.~\eqref{math:densityp} 
identifies a family of profiles in which $\beta$ and $\gamma$ determine 
the dependence of the profile at large and small 
distances, respectively, with the sharpness of the 
transition given by $\alpha$ \cite{Graham:2005xx}. 
The slope of the distribution changes at a characteristic 
spatial scale determined by $a_0$, with $\rho_0$ being 
the corresponding density.
Here we focus on two models, commonly used to 
interpret DM distribution emerging from numerical 
simulations and astrophysical observations: (i) 
the Hernquist profile corresponding to 
$(\alpha,\beta,\gamma)=(1,4,1)$ \cite{Hernquist:1990be} and, 
(ii) the Navarro-Frenk-White (NFW) distribution 
obtained by fixing $(\alpha,\beta,\gamma)=(1,3,1)$ \cite{Navarro:1996gj}. 
It is known that the NFW model predicts a mass function which 
diverges logarithmically with $r$. For this reason we prescribe 
a radial cut-off $r_c$, such that $M_\tn{halo}(r>r_c)=0$. We 
will analyse how this parameter affects the geodesic 
properties of the spacetime, and the emitted axial fluxes. 
Along with profiles inspired by Eq.~\eqref{math:densityp}, we 
also exploit our framework to study the $1/r^n$ Einasto 
model \cite{1969Afz.....5..137E}: 
\begin{equation}
\rho(r)=\rho_e \tn{exp}
\left\{-d_n[(r/r_e)^{1/n}-1]\right\}\ ,
\end{equation}
with $n=6$, $d_n\sim17.67$ \cite{Prada:2005mx,Graham:2005xx}, 
and $\rho_e$ being the density at 
the radius $r_e$ that defines a 
volume containing half of the halo mass, 
fixed hereafter to $r_e=a_0$. 

Newtonian and relativistic analyses 
show that density profiles with a BH sitting at their 
core, vanish at the horizon, and develop a 
cusp with a lengthscale dictated by 
the BH mass \cite{Gondolo:1999ef,Sadeghian:2013laa}. 
The details of the overdensity depends on the 
specific form of the profile, and may be relevant 
for an accurate modelling of the GW signals 
emitted by coalescing binaries \cite{Speeney:2022ryg}.
However, for the purpose of this work we encode the 
features of the accretion growth by re-scaling the 
density profile according to 
$\rho(r)\rightarrow \rho(r)(1-2M_\tn{BH}/r)$, 
following the results of \cite{Cardoso:2020iji}.
A full numerical treatment of 
the DM accretion onto the BH will be discussed 
in a forthcoming work, in which we will also 
extend current calculations to integrate 
polar perturbations for generic density 
profiles. 
Figure~\ref{fig:profiles} shows the behavior of 
$\rho(r)$ for the three cases described above, 
for prototype choices of $(M_\tn{halo},a_0,r_c,r_e)$.

\section{Numerical procedure}
\begin{figure}[h]
    \centering
    \includegraphics[width=8cm]{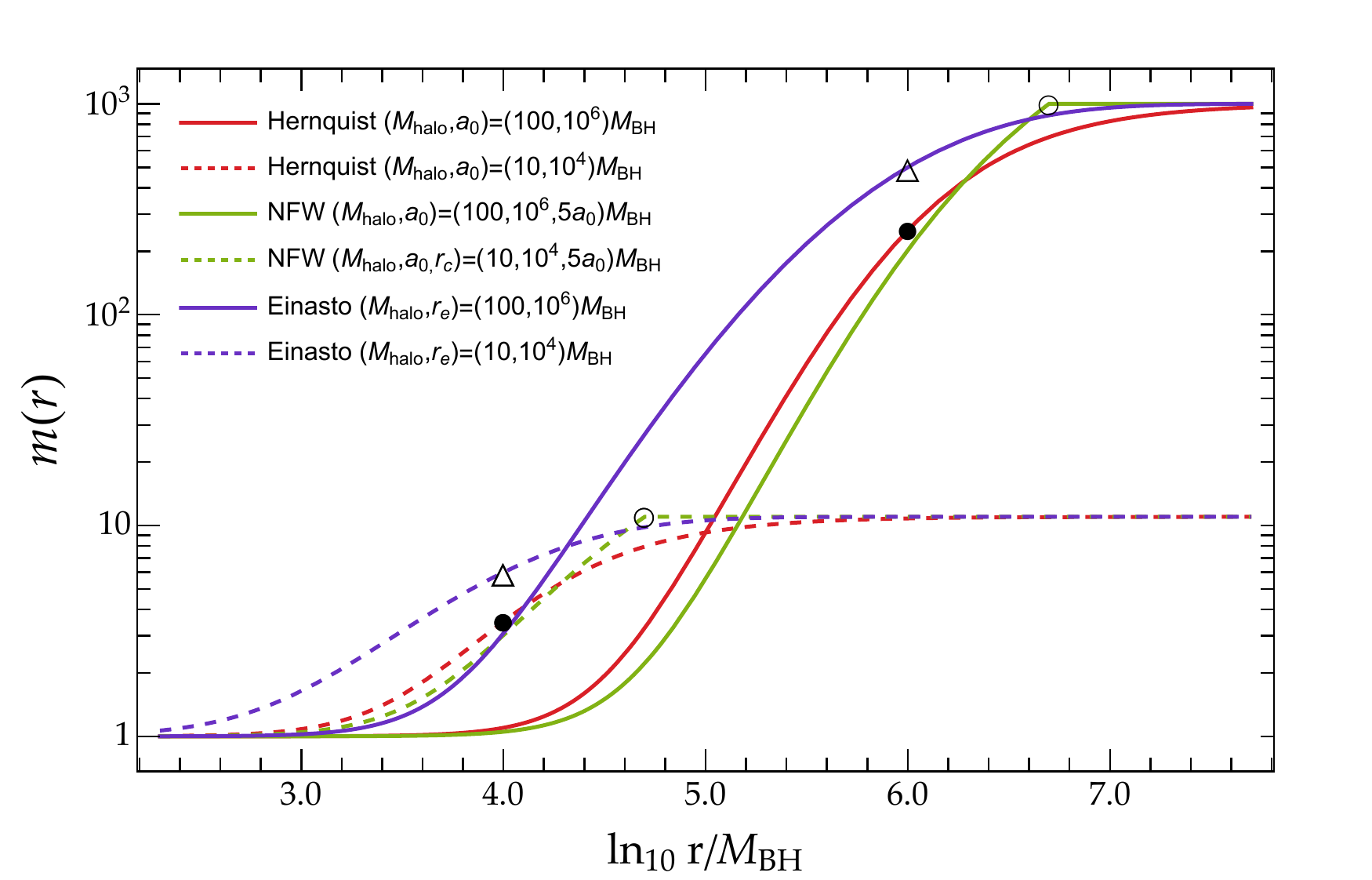}
    \includegraphics[width=8cm]{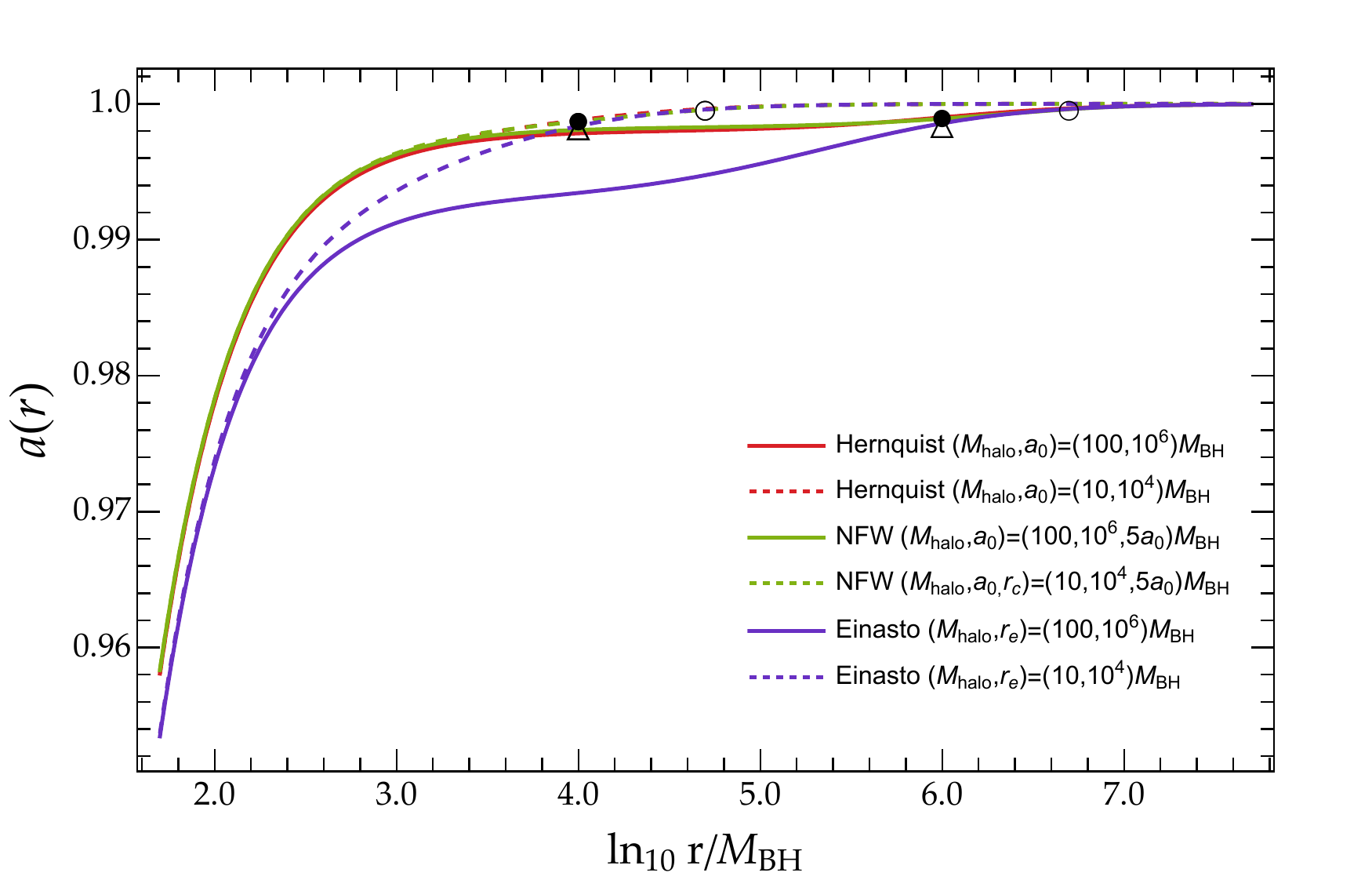}
    \caption{Metric components $m(r)$ and $a(r)$ as a 
    function of the coordinate radius, for different 
    profiles and configurations. Black dots, empty 
    circles and empty triangles along curves identify 
    the value of $a_0$, $r_c$ and $r_e$ for the Hernquist, 
    NFW and Einasto model, respectively.}
    \label{fig:backround}
\end{figure}
We have solved the background and the perturbation 
equations according to the following numerical 
procedure:
\begin{enumerate}
    \item We start by choosing a density profile 
    $\rho(r)$ according to the Hernquist, the NFW or the 
    Einasto model. We integrate the equation for the mass 
    function $m(r)$ from the horizon $r_\tn{h}=2M_\tn{BH}$, 
    where $m(r_h)=M_\tn{BH}$, to a coordinate radius $r_\tn{out}$ 
    that corresponds to our spatial infinity, and which 
    guarantees asymptotic flatness. 
    We take $r_\tn{out} \gtrsim 10^7 a_0$. 
    Changing 
    this value by more than three orders of magnitude 
    does not affect our results. We then solve backward the 
    equation for the metric function $a(r)$ assuming as initial 
    condition in the far field limit that $m(r\rightarrow r_\tn{out})=M_\tn{BH}+M_\tn{halo}$ and
    \begin{equation}
    a(r)=1-\frac{2(M_\tn{BH}+M_\tn{halo})}{r}+\mathcal{O}\left(1/r^3\right)\ . 
    \end{equation}
    Note that $M_\tn{halo}$ is therefore the total environmental mass outside the BH. As we explained, although geared towards DM distributions our results and techniques are applicable to any spherically symmetric environment. Examples of metric functions $a(r)$ and $m(r)$ are shown in Fig.~\ref{fig:backround}.
    \item The numerical solution for $m(r)$ and $a(r)$ allows to 
    compute the tangential pressure $P_t(r)$, as well as the geodesic 
    quantities, like the ISCO and the light ring frequencies, 
    $\Omega_\tn{ISCO}$ and $\Omega_\tn{LR}$.
    \item We integrate the master equation \eqref{math:masterax} 
    for $\psi_{\ell m}$ using a standard Green function approach \cite{Cardoso:2021wlq}. We first solve the associated homogeneous 
    problem requiring that the physical solution satisfies 
    pure ingoing/outgoing wave boundary condition at the horizon and 
    at infinity, namely:
    \begin{align}
    \psi_{\ell m}^{(\tn{in})}&= e^{-i \omega r_*} \sum_{i=0}^{n_\tn{in}} \alpha_i (r-r_\tn{h})^i\ , \nonumber\\
    \psi_{\ell m}^{(\tn{out})}&=e^{+i \omega r_*} \sum_{i=0}^{n_\tn{out}} \frac{\beta_i}{r^i}\ ,
    \end{align}
    where we fix\footnote{We verified that 
    larger values of $n_\tn{in}$ and $n_\tn{out}$ 
    do not change our results to the precision required here.} $n_\tn{in}=n_\tn{out}=5$.
    The coefficients $(a_i,b_i)$  are obtained by 
    solving the homogeneous equation at each order 
    in $(r - r_\tn{h})$ and $1/r$, and setting 
    $\alpha_0=\beta_0=1$. To this aim we also need to expand 
    the metric functions $m(r)$ and $a(r)$ at both boundaries. 
    At the horizon we consider the following ansatz:
    \begin{align}
    m(r) &= M_\tn{BH} + 
    \sum_{i=1}^{n_\tn{in}}m^{(i)}(r_\tn{h})(r-r_\tn{h})^i  \ , \\
    a(r) &= \sum_{i=1}^{n_\tn{in}} a^{(i)}(r_\tn{h})(r-r_\tn{h})^i  \ ,
    \end{align}
    where the coefficients are found numerically 
    using the interpolated numerical solutions for 
    $m(r)$ and $a(r)$ found at step 1.
    At infinity we assume consistently that 
    $m(r_\tn{out})= M_\tn{BH} + M_\tn{halo}$
    and $a(r_\tn{out}) = 1 - 2(M_\tn{BH} + M_\tn{halo})/r_\tn{out}$.
    The full solution at infinity is 
    then obtained integrating the homogeneous 
    component over the source term:
    \begin{align}
    \psi_{\ell m}^{\tn{out}}=&\lim_{r_\star\rightarrow r_\tn{out}}\psi_{\ell m}(r_\star)\nonumber\\
    =&e^{+i \omega r_\star} 
    \int_{r_\tn{h}}^{r_\tn{out}}\frac{\psi_{\ell m}^{(r_\tn{h})}S_{\ell m}}{W}dr_\star\ ,\label{math:nhomgsol}
    \end{align}
    where $W=d\psi_{\ell m}^{(\tn{out})}/dr_\star\psi_{\ell m}^{(\tn{in})}-d\psi_{\ell m}^{(\tn{in})}/dr_\star\psi_{\ell m}^{(\tn{out})}$ is the Wronskian.
    \item From Eq.~\eqref{math:nhomgsol} we obtain for 
    each multipole component $(\ell,m)$ the GW flux at infinity:
    \begin{equation}
    \dot{E}^\infty_{\ell m} = \frac{1}{16\pi} \frac{(\ell+2)!}{(\ell-2)!} \abs{\psi_{\ell m}}^2 \quad (\ell+m \quad \tn{odd})\ .
    \end{equation}
\end{enumerate}
The codes developed to integrate the background 
and the first order equations are freely available 
as a \texttt{Mathematica} package at \cite{SGREP_REPO}. 

\section{Results}\label{sec:results}

\subsection{Geodesic properties}\label{sec:results_geo}
\begin{figure*}[h]
    \centering
    \includegraphics[width=15cm]{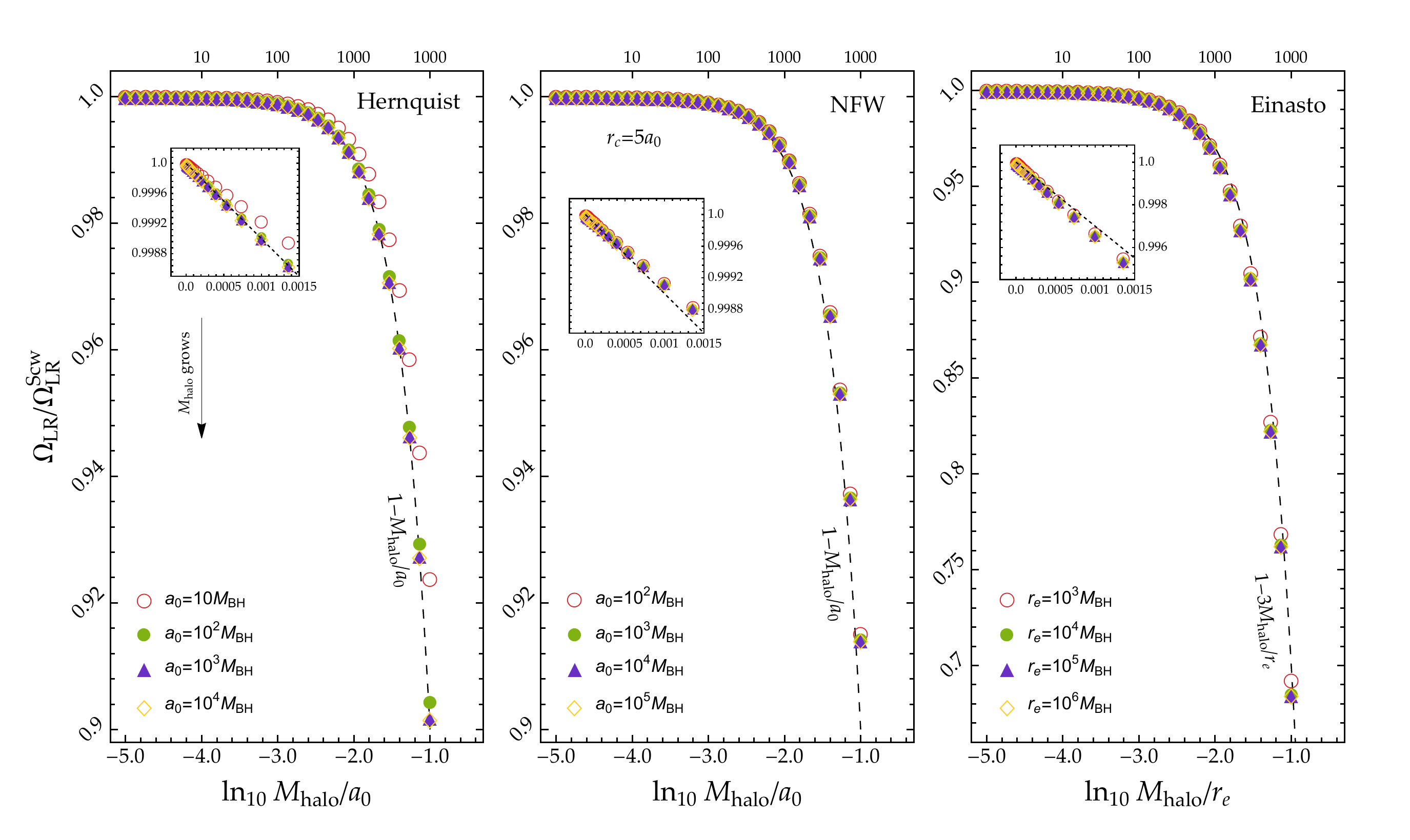}\\
    \includegraphics[width=15cm]{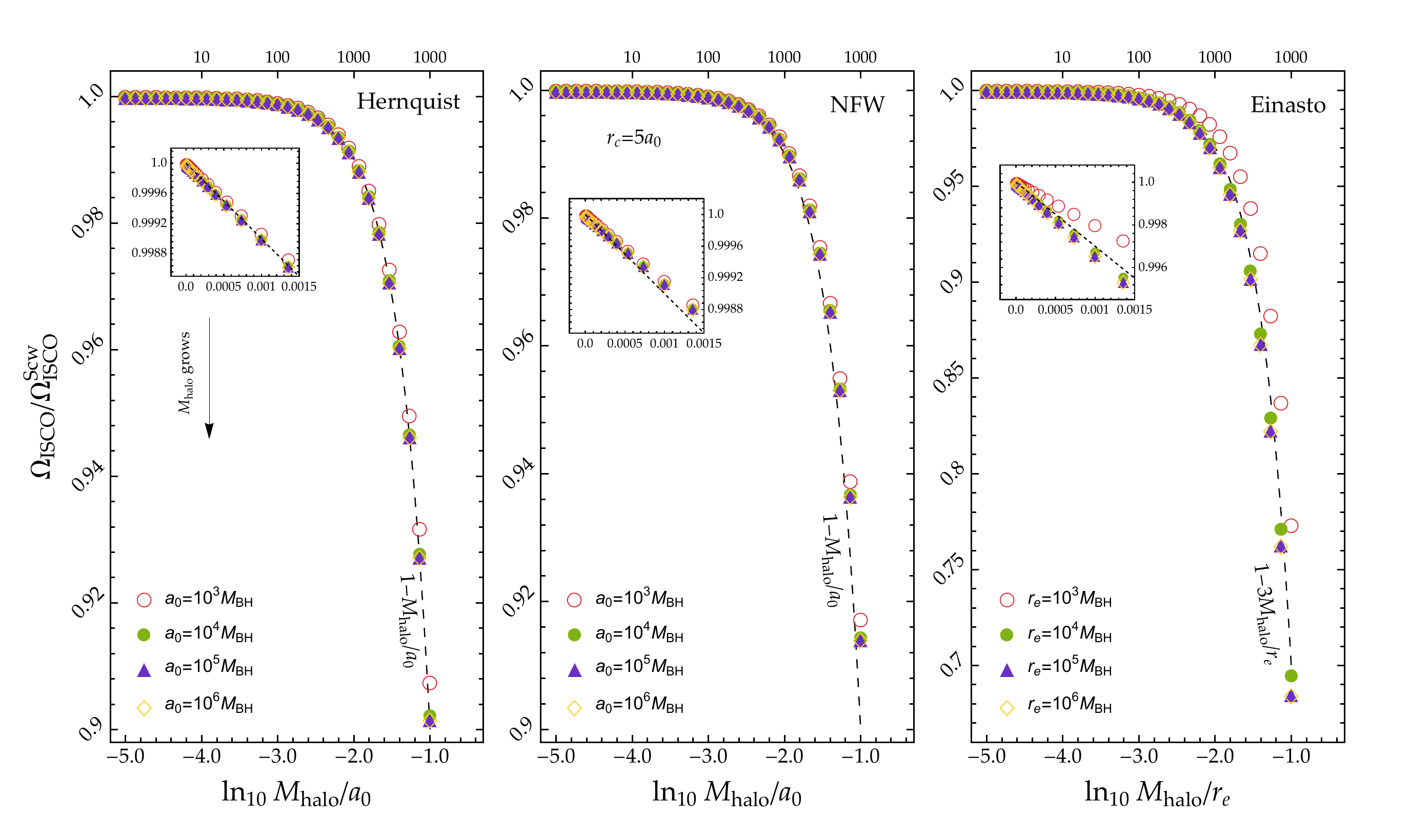}\\
    \caption{{\bf Top row:} Light ring frequencies as a 
    function of the redshift parameter $M_{\rm halo}/a_0$ for the Hernquist and NFW profiles, and $M_{\rm halo}/r_e$ for the Einasto model. Frequencies are normalized to the Schwarzschild value $\Omega_\tn{LR}^\tn{Scw}=1/3\sqrt{3} M_\tn{BH}$. 
    Different dots and colors refer to various choices 
    of $a_0$ and $r_e$. The arrow in the first panel 
    on the left identifies the direction in which, for a 
    given  $M_\tn{halo}/a_0$ (and $M_\tn{halo}/r_e)$, $M_\tn{halo}$ grows. 
    The top axis show the values of the 
    halo mass for a reference $a_0=10^5M_\tn{BH}$ 
    ($r_e=10^5M_\tn{BH}$). 
    The black dashed line corresponds to a re-scaling 
    of the frequencies 
    $\Omega_\tn{LR}=\Omega_\tn{LR}^\tn{Scw}(1-M_\tn{halo}/a_0)$ for 
    Hernquist and NFW, and $\Omega_\tn{LR}=\Omega_\tn{LR}^\tn{Scw}(1-3M_\tn{halo}/r_e)$ for Einasto. 
    For the NFW profile we show values of the 
    frequencies for $r_c=5a_0$. 
    The inset in each panel shows a zoom 
    on the low $M_\tn{halo}/a_0$ and $M_\tn{halo}/r_e$ 
    regime. {\bf Bottom row:} Same as top but for the orbital frequencies at 
    the Innermost Stable Circular Orbits, with 
    $\Omega_\tn{ISCO}^\tn{Scw}=1/6\sqrt{6}M_\tn{BH}$.}
    \label{fig:freq_geo}
\end{figure*}
Our numerical approach allows us to study, 
along with the axial perturbations, the geodesic 
properties of the background.

The orbital frequencies at the light-ring and at the 
ISCO, are key to determine the observational 
signatures of bodies and radiation surrounding the BH \cite{EventHorizonTelescope:2019dse,2018AA...618L..10G}. 
The frequency, as measured by far-away observers, is shown in the top row of Figure~\ref{fig:freq_geo}, $\Omega_\tn{LR}$, 
normalized to the Schwarzschild value 
$\Omega_\tn{LR}^\tn{Scw}=1/(3\sqrt{3}M_\tn{BH})$ for the 
Hernquist and the NFW model, as a function of   
$M_\tn{halo}/a_0$, for different values 
of $a_0$. We fix $r_c=5a_0$ for the NFW profile. 
Results for the Hernquist case provide a fully 
numerical confirmation of the analysis presented 
in Ref.~\cite{Cardoso:2021wlq}. 
Changes with respect to the Schwarzschild solution 
can be interpreted in terms of a redshift 
scaling of the frequencies 
$\Omega_\tn{LR}/\Omega_\tn{LR}^\tn{Scw}\sim1-z$, 
with $z=M_\tn{halo}/a_0$. Such dependence becomes increasingly 
more accurate as $z$ decreases, and for larger values 
of $a_0$, as shown in the inset of the panels, which 
provides a zoom on the low-$z$ regime.

\begin{figure}[h]
    \centering
    \includegraphics[width=9cm]{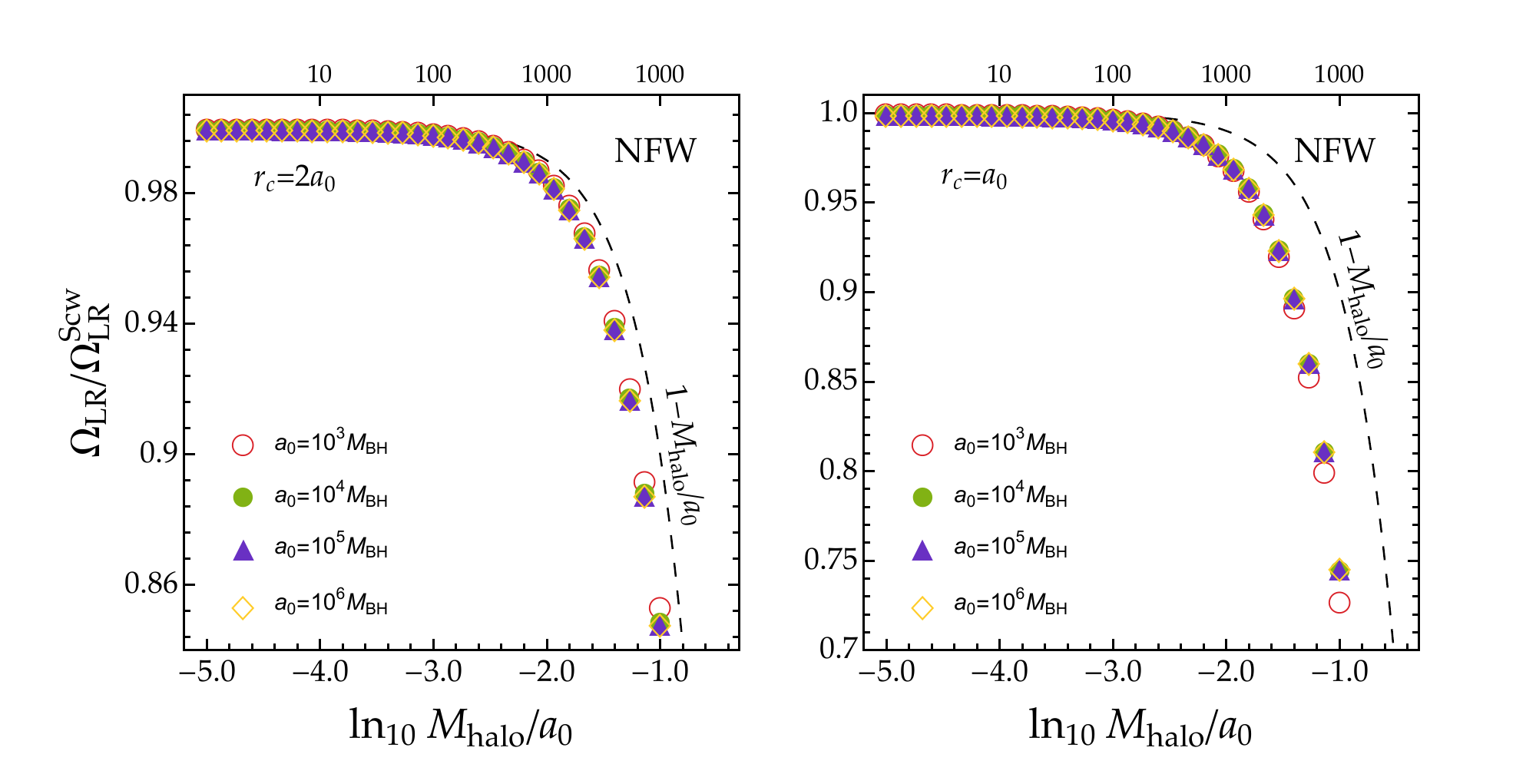}\\
    \includegraphics[width=9cm]{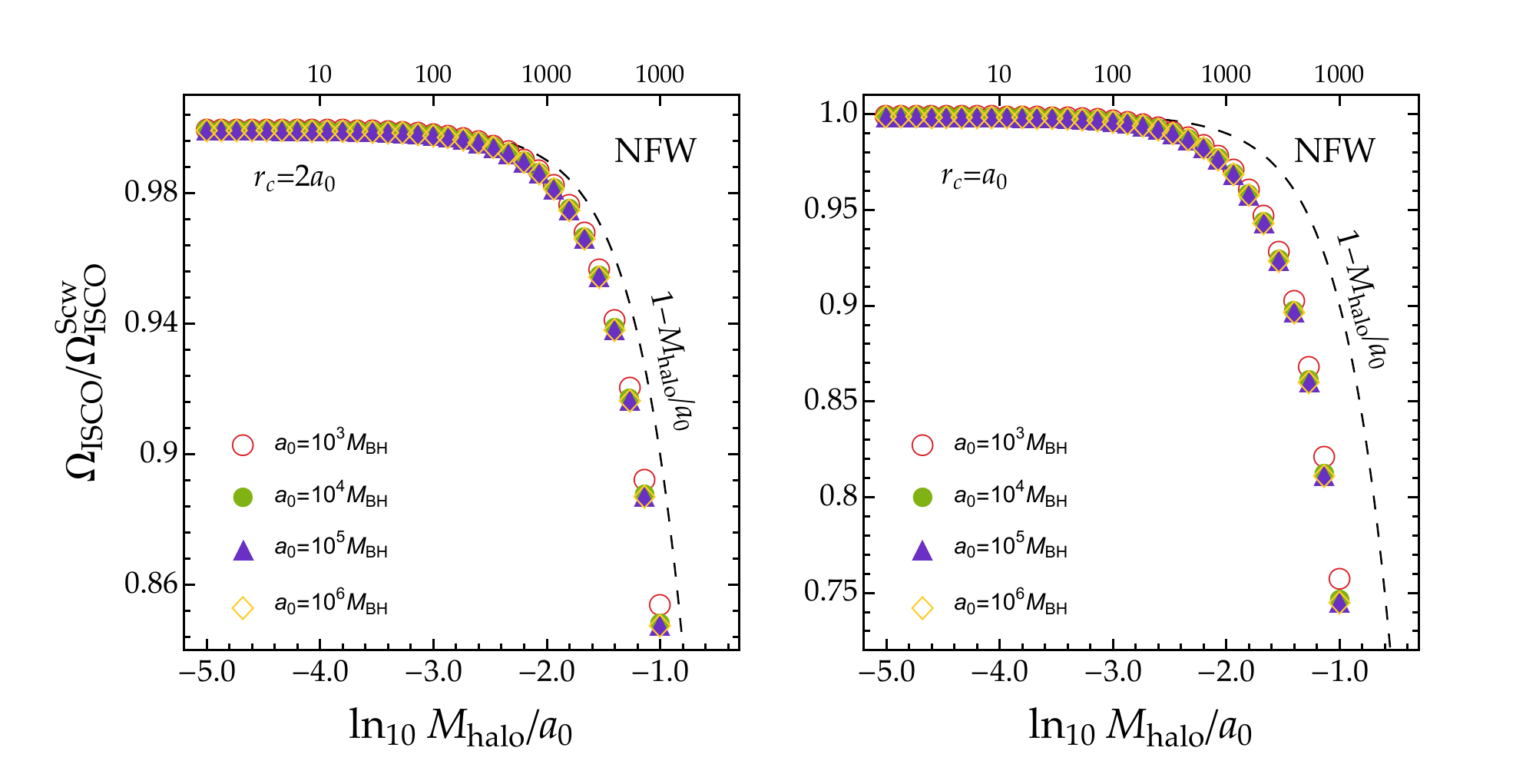}\\
    \caption{Light ring (top) and ISCO (bottom) 
    frequencies for the NFW  density profile as a 
    function of $M_{\rm halo}/a_0$, normalised 
    to the corresponding Schwarzschild values, for 
    different $a_0$. Left and right panels correspond 
    to a cut-off radius for the matter distribution 
    of $r_c=2a_0$ and $r_c=a_0$. The Top axis identify 
    the halo mass $M_\tn{halo}/M_\tn{BH}$ 
    for $a_0=10^5M_\tn{BH}$.}
    \label{fig:app_freq_geo}
\end{figure}

Similar considerations hold for the ISCO frequencies.
We have fitted our numerical data for 
$a_0\in[10^2,10^7]M_\tn{BH}$ and $z\leq0.001$ as   
\begin{equation}
\frac{\Omega_\tn{LR,ISCO}}{\Omega_\tn{LR,ISCO}^\tn{Scw}}=
c_0+c_1 z+c_2z^2+c_3 M_\tn{halo}/a_0^2\ ,
\end{equation}
recovering for the coefficients 
$c_i$ the values found in closed form 
in Ref.~\cite{Cardoso:2021wlq}, through a small $z$ 
expansion of $\Omega_\tn{LR,ISCO}$. For $z\lesssim 0.01$, 
a simple scaling $\Omega_\tn{LR,ISCO}=\Omega_\tn{LR,ISCO}^\tn{Scw}(1-z)$ is able to reproduce the data 
with a relative errors smaller than $1\%$.
This behavior also seems to apply to the  NFW case, 
shown in the center panel of Fig.~\ref{fig:freq_geo} 
assuming $r_c=a_0$. 
While the slope of the scaling depends on the 
actual value of $r_c$, for $z$ smaller than 
$10^{-3}$, we recover an universal trend. 
Results for $r_c=a_0$ and $r_c=2a_0$ are drawn 
in Fig.~\ref{fig:app_freq_geo}. 
Regardless of the specific choice of the cutoff, 
changes with respect to the vacuum case follow a 
$M_\tn{halo}/a_0$ scaling. We have explored different configurations 
within $r_c\in[1,5]a_0$ and $a_0\in[10^2,10^7]M_\tn{BH}$. 
In this domain, for $M_\tn{halo}/a_0\le 0.01$ and 
$M_\tn{halo}>M_\tn{BH}$ we find that the 
semi-analytic fits  
\begin{equation}
\frac{\Omega_\tn{LR}}{\Omega_\tn{LR}^\tn{Scw}}=1-\frac{p_1 z}{(r_c/a_0)^{p_2}}\ ,\ 
\quad\ \frac{\Omega_\tn{ISCO}}{\Omega_\tn{ISCO}^\tn{Scw}}=1-\frac{s_1z}{(r_c/a_0)^{s_2}}\ ,
\end{equation}
where $(p_1,p_2,s_1,s_2)=(2.546,0.698,2.535,0.6956)$, 
are able to describe our numerical results 
with relative accuracy better than $\sim 0.1\%$.

Finally, the right panels of Fig.~\ref{fig:freq_geo} show 
the geodesics analysis for the Einasto model. 
In this case deviations from Scwarzschild for both 
the light ring and ISCO frequencies highlight 
a dependence on $M_\tn{halo}/r_e$. 
As $r_e$ increases $\Omega_\tn{LR,ISCO}/\Omega_\tn{LR,ISCO}^\tn{Scw}$ 
follow a scaling $\sim 1-3M_\tn{halo}/r_e$, which 
becomes independent from $r_e$ as 
$M_\tn{halo}/r_e$ decreases. 

\begin{table}[h!]
\centering
    \begin{tabular}{ccccccc}
         \hline \hline
         $\ell$ & $\abs{m}$ & Schw & Hernquist & \begin{tabular}{@{}c@{}} NFW \\ $r_c = a_0$ 
         \end{tabular} & \begin{tabular}{@{}c@{}} NFW \\ $r_c =5 a_0$ 
         \end{tabular} &  \begin{tabular}{@{}c@{}} Einasto \\ $r_e = a_0$ 
         \end{tabular}  \\
         \hline \hline
         2 & 1 & 7.8175e-7 & 7.8019e-7 & 7.7771e-7 & 7.8039e-7 & 7.7642e-7 \\ 
         \hline
         3 & 2 & 2.3974e-7 & 2.3926e-7 & 2.3850e-7 & 2.3933e-7 & 2.3811e-7 \\ 
         \hline
         4 & 1 & 7.9715e-13 & 7.9555e-13 & 7.9301e-13 & 7.9576e-13 & 7.9158e-13 \\
         \hline
         4 & 3 & 5.4572e-8 & 5.4462e-8 & 5.4290e-8 & 5.4477e-8 & 5.4200e-8 \\
         \hline \hline 
    \end{tabular}
    \caption{Axial fluxes (normalized to $m^2_p/M^2_\tn{BH}$) for 
    a secondary sitting at $r_p=8 M_\tn{BH}$. 
    We focus on halo configurations with $M_{\rm halo}/a_0=10^{-3}$ and 
    $a_0=10^5 M_{\rm BH}$. For the Einasto profile we assume $r_e=a_0$.}
    \label{table1}
\end{table}
\subsection{Gravitational-wave emission}\label{sec:results_GW}
\begin{figure}[h]
    \centering
    \includegraphics[width=9.2cm]{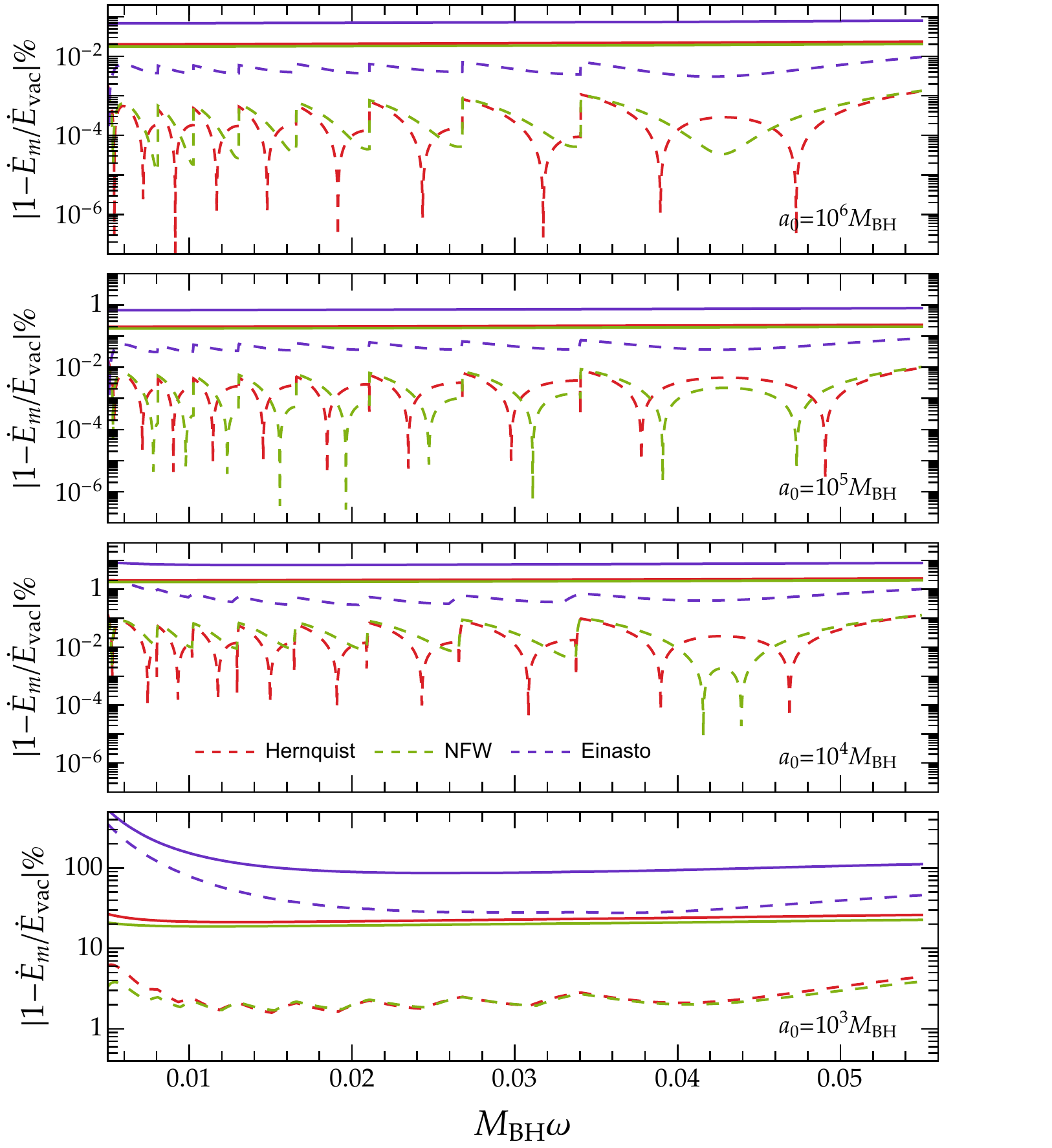}
    \caption{Relative percentage difference between the 
    matter and vacuum axial $\ell=2$ $m=1$ flux emitted 
    by an EMRI, as a function of the GW frequency. 
    We assume $M_\tn{halo}=100 M_\tn{BH}$ for the 
    halo configuration. Fluxes for the NFW and the 
    Einasto model are obtained fixing $r_c=5a_0$ 
    and $r_e=a_0$, respectively. Dashed (solid) 
    lines correspond to relative differences with 
    respect vacuum redshifted (unredshifted) results.}
    \label{fig:redshift}
\end{figure}

We can now focus on the axial GW emission. 
Tables~\ref{table1}-\ref{table2} show the values of 
the energy released at infinity for different mode 
configurations, compared against the vacuum components. 
We computed the latter through the Black Hole 
Perturbation Toolkit \cite{BHPToolkit}.
For a given BH mass, GW fluxes are smaller in 
the case of non-vacuum environments, and 
decrease as the compactness of the halo, 
either, $M_\tn{halo}/a_0$ or $M_\tn{halo}/r_e$, 
grows. For the NFW model this behavior also 
depends on the cutoff radius, with 
$\dot{E}^{\infty}_{\ell m}$ becoming smaller 
as $r_c$ shrinks.
As shown in \cite{Cardoso:2021wlq} however, 
differences between the Schwarzschild and the halo 
case can be interpreted, for the axial sector, 
in terms of a redshift effect. Our results 
support this picture across different halo models. 
Figure~\ref{fig:redshift} shows indeed 
the relative difference between matter and vacuum 
fluxes as a function of the GW frequency, for the 
$(\ell,m)=(2,1)$ mode, and $M_\tn{halo}=10^2M_\tn{BH}$. 
Dashed (solid) curves correspond to redhifted 
(unredshifted) fluxes in vacuum, obtained 
by scaling
\begin{equation}
\Omega^\tn{vac}\rightarrow \Omega/\gamma\ \ ,\ 
m_p^\tn{vac}\rightarrow m_p\gamma\ \ ,\ 
\omega^\tn{vac}\rightarrow \omega/\gamma\ ,
\end{equation}
where $\gamma=1-\delta M_\tn{halo}/a_0$ for the 
Hernquiest and NFW profiles and $\gamma=1-\delta M_\tn{halo}/r_e$ 
for the Einasto model.
In agreement with the geodesic analysis, we find 
the best match between the Schwarzschild 
and halo fluxes assuming $\delta=1$, $\delta\sim 0.9$ 
in the first two cases, and $\delta \sim 3$ for  
the Einasto distribution. The latter yields in 
general the largest differences with respect to 
the vacuum evolution. The agreement between the 
redshift and matter results improves again for less 
dense halos, and deteriorates only for very compact 
(and unrealistic) configurations with $M_\tn{halo}/a_0=0.1$.

\begin{table}[h!]
\centering
    \begin{tabular}{ccccccc}
         \hline \hline
         $\ell$ & $\abs{m}$ & Schw & Hernquist & 
         \begin{tabular}{@{}c@{}} NFW \\ $r_c = a_0$ \end{tabular} & \begin{tabular}{@{}c@{}} NFW \\ $r_c =5 a_0$ \end{tabular} &  \begin{tabular}{@{}c@{}} Einasto \\ $r_e = a_0$ \end{tabular}  \\
         \hline \hline
         2 & 1 & 7.8175e-7 & 6.3598e-7 & 4.3470e-7 & 6.5338e-7 & 4.1168e-7 \\ 
         \hline
         3 & 2 & 2.3974e-7 & 1.9545e-7 & 1.3408e-7 & 2.0060e-7 & 1.3085e-7 \\ 
         \hline
         4 & 1 & 7.9715e-13 & 6.0233e-13 & 3.6128e-13 & 6.4033e-13 & 1.6413e-13 \\
         \hline
         4 & 3 & 5.4572e-8 & 4.4570e-8 & 3.0668e-8 & 4.5706e-8 & 3.0968e-8 \\
         \hline \hline
         
    \end{tabular}
    \caption{Same as Table~\ref{table1} but for an halo with 
    $M_{\rm halo}/a_0=10^{-1}$.}
    \label{table2}
\end{table}

\section{Conclusions}

Our work provides a step forward towards a general relativistic 
description of compact sources evolving within non-trivial environments. 
We have extended the analysis carried out in Ref.~\cite{Cardoso:2021wlq,Cardoso:2022whc}, developing a numerical 
approach to find BH solutions embedded in an anisotropic 
fluid with a generic density profile. We have exploited this 
approach to study the impact of different DM 
distributions on the BH spacetime, and on the 
evolution of very asymmetric binaries, with small mass ratios. 

While changes with respect to vacuum backgrounds depend on 
the specific setup, our results confirm the existence of 
features common to different models. Deviations from 
Schwarzschild predictions scale with the halo density 
and increase in the presence of overdensities 
close to the BH horizon, as spikes induced by 
matter accretion.
Such changes however can be interpreted, and quantified, in terms of redshift 
of the geodesic properties and of the GW frequencies. For halo 
models belonging to the two-parameter family described by 
Eq.~\eqref{math:densityp}, like the Hernquist and the NFW 
profiles, the redshift scaling approaches a universal 
behavior when the halo compactness $M_\tn{halo}/a_0$ becomes 
smaller than $10^{-3}$. Similar considerations hold for the 
Einasto distribution, for which redshift of frequencies and 
fluxes is dictated by the spatial scale of the profile. 

While the approach we have developed is completely general, 
our numerical results assume a simple prescription to 
describe the DM accretion onto the BH. 
We have explored the dependence of our 
conclusions on such assumption by adopting a second 
scaling of the halo profile, 
$\rho(r)\rightarrow \rho(r)(1-4M_\tn{BH}/r)$, 
which mimics the correct behavior found by fully 
relativistic calculations of adiabatic accretion 
\cite{Sadeghian:2013laa}. We have studied the 
properties of the geodesic motion and of the axial 
emission, finding for both a qualitative agreement with 
the analysis discussed in Sec.~\ref{sec:results}. 

The details of the matter distribution, however, 
can be relevant to accurately determine the actual 
evolution of binaries throughout the coalescence, and 
build waveform models for next generation of detectors 
\cite{Speeney:2022ryg}. We are working to include 
the full general relativity treatment of the 
the DM spike distribution within the formalism 
devised in this work, further extending the set of 
gravitational perturbations to the polar sector. 
The results of such analysis will be presented in a 
forthcoming paper with a full adiabatic evolution 
of EMRI in non-vacuum environments.

\noindent{{\bf{\em Acknowledgments.}}}
E.F. acknowledges financial support from ENS de Lyon.
This work makes use of the Black Hole Perturbation Toolkit.
V.C.\ is a Villum Investigator and a DNRF Chair, supported by VILLUM Foundation (grant no. VIL37766) and the DNRF Chair program (grant no. DNRF162) by the Danish National Research Foundation. V.C.\ acknowledges financial support provided under the European Union's H2020 ERC Advanced Grant ``Black holes: gravitational engines of discovery'' grant agreement
no.\ Gravitas--101052587. Views and opinions expressed are however those of the author only and do not necessarily reflect those of the European Union or the European Research Council. Neither the European Union nor the granting authority can be held responsible for them.
This project has received funding from the European Union's Horizon 2020 research and innovation programme under the Marie Sklodowska-Curie grant agreement No 101007855.
We thank FCT for financial support through Projects~No.~UIDB/00099/2020 and UIDB/04459/2020.
We acknowledge financial support provided by FCT/Portugal through grants 
2022.01324.PTDC, PTDC/FIS-AST/7002/2020, UIDB/00099/2020 and UIDB/04459/2020.
\bibliography{biblio}

\end{document}